# Anatomy of a Historic Blackout: Decoding Spatiotemporal Dynamics of Power Outages and Disparities During Hurricane Beryl

Xiangpeng Li[1*], Junwei Ma[1], Ali Mostafavi[2]

[1] Ph.D. student. Urban Resilience.AI Lab, Zachry Department of Civil and Environmental Engineering, Texas A&M University, College Station, Texas, United States.

[2] Professor. Urban Resilience.AI Lab, Zachry Department of Civil and Environmental Engineering, Texas A&M University, College Station, Texas, United States.

[*] Corresponding author: Xiangpeng Li, E-mail: xplli@tamu.edu.

**Abstract:**

Power outages are a significant consequence of natural hazards, severely disrupting communities' restoration and recovery processes. Despite the increasing frequency and impact of hazard-induced power outages, empirical studies examining their spatial and temporal characteristics across impacted regions remain limited. This dearth of empirical insights inhibits the ability to quantify impacts and examine vulnerability and equity issues for effective resilience investments. This study investigates the spatial patterns and temporal variations in outage duration, intensity, and restoration/recovery following the 2024 Hurricane Beryl in Houston, Texas. This historic blackout caused widespread power disruptions across the Houston metropolitan area, leaving more than 2 million customers without power over several days, resulting in more than 143 million total customer-out hours. By examining the dynamic interplay between outage impact, recovery features, and socioeconomic and infrastructural factors, the analysis identified key determinants contributing to disparities in power outage impacts and recovery efficiency delineated by Zip Code across Houston. The findings reveal that areas with higher population density and proximity to the hurricane's path experienced more severe initial impacts. Regions with higher median income showed faster recovery, while lower-income areas exhibited prolonged restoration periods, even with favorable infrastructural conditions, suggesting disparities in restoration speed. The study also highlights how urban development features, such as road density and land elevation, explain spatial disparities in power outage impacts and recovery. This research advances the understanding of power outage dynamics in large metropolitan regions through four key contributions: (1) empirical characterization of outages from a historic hurricane, highlighting infrastructure vulnerabilities in a high-density urban context; (2) comprehensive analysis using multiple metrics to capture spatiotemporal dynamics of outages and restoration; (3) leveraging of high-resolution outage data at fine geographic scales and frequent intervals to quantify and reveal previously masked spatial disparities; and (4) systematic examination of socioeconomic, urban development, and environmental factors in shaping disparities in outage impacts and recovery timelines. These findings provide infrastructure managers, operators, utilities, and decision-makers with crucial empirical insights to quantify power outage impacts, justify resilience investments, and address vulnerability and equity issues in the power infrastructure during hazard events.

**Keywords:** power outages, spatial analysis, community recovery, infrastructure resilience, equity.

## 1. Introduction

Power outages represent a significant consequence of natural disasters, especially in highly populated urban areas , where electricity is essential for daily life, emergency services, and critical infrastructure systems (Flores, McBrien et al. 2023, Xu, Feng et al. 2024, Zhou, Hu et al. 2024). With the increasing frequency and intensity of severe weather events, understanding the extent to which power systems are vulnerable to these disturbances is crucial to inform mitigation and response plans and actions (Baik, Davis et al. 2020, Feng, Ouyang et al. 2022, Do, McBrien et al. 2023). Of particular importance is the empirical characterization of the spatial and temporal patterns of outages and restoration speed. Empirical examinations of the extent and speed of power restoration are critically important for both scientific inquiry and practical decision-making (Vaiman, Chen et al. 2011). From a scientific perspective, systematic data on the spatial and temporal characteristics of outages enables researchers to identify patterns, test hypotheses about underlying causes, and model the complex interplay of infrastructure vulnerabilities and environmental stressors. Such insights are especially valuable in refining resilience theories and guiding further research into equitable energy access. On a practical level, robust empirical findings inform utility providers, policymakers, and emergency management officials in devising targeted solutions to improve restoration strategies, prioritize resources effectively, and minimize societal and economic losses (Esparza, Li et al. , Li, Ma et al. 2024, Ma, Li et al. 2024). In contexts where limited empirical studies exist, the ability to quantify impacts and evaluate vulnerability and equity issues is  hampered, hindering both immediate relief efforts and longer-term resilience planning (Li and Mostafavi 2022, Li and Mostafavi 2024). In the absence of clear, data-driven insights, decision-makers cannot accurately pinpoint where restoration efforts are lagging, how resources should be allocated, or which communities face disproportionate burdens (Coleman, Li et al. 2023). This dearth of robust empirical studies on how outages unfold and are resolved across diverse geographic contexts also poses significant challenges for both research and practice. This blind spot in understanding perpetuates inequities and hampers targeted interventions to strengthen infrastructure, particularly in regions that are socioeconomically vulnerable or geographically isolated. Furthermore, the inability to capture and learn from historical patterns and trends limits the development of predictive models, ultimately undermining both system reliability and public trust (Habbal, Ali et al. 2024). Over time, these gaps erode the collective capacity to mitigate risks, respond effectively to crises, and ensure equitable access to electricity during hazard events.

Recognizing the importance of empirical studies related to power outages during hazard events, a number of recent studies have focused on examining the characteristics of outages For instance, Coleman, Esmalian et al. 2023 conducted a detailed analysis of outage patterns during the Winter Storm Uri (2021) and Hurricane Ida (2021), identifying significant disparities in outage durations across different sociodemographic groups. Similarly, (Li and Mostafavi 2024) examined more than two hundred power-grid resilience curves related to power outages in three major extreme weather events in the United States (2023 Austin Ice Storm, 2017 Hurricane Irma, and 2021 Hurricane Ida) to identify two primary archetypes for power grid resilience curves. (Best, Kerr et al. 2023) investigated the power outages during 2012 Hurricane Isaac and found that infrastructure damage and recovery times resulting from hurricanes disproportionately affect socioeconomically vulnerable populations and racial minorities. (Jamal and Hasan 2023) used changes in the number of Facebook users during Hurricane Ida to understand transient loss in community resilience in Louisiana. These studies not only contribute to a growing body of literature that forms the foundation of empirical assessments of power outage vulnerability and resilience, paving the way for more targeted interventions and policy decisions, but also highlights the complex interplay between environmental stressors, infrastructure resilience, and social vulnerability in the context of power outages.

Despite the growing recognition of the importance of empirical studies on power outages, these studies are still rather limited. The main limitations of the existing empirical studies are twofold. First, existing studies focus on limited aspects of power outages and restoration using limited outage metrics, thus hindering the ability to fully quantify the severity and to characterize spatiotemporal variations across different regions of impacted communities. Second, the existing empirical studies lack the spatial and temporal resolution needed to specify various outage-related features and to analyze variations in outage features across different regions of affected communities. Multiple studies rely on large geographic units, such as counties or census tracts, which lack the resolution to capture localized variations in outage impacts and social vulnerability (Esparza, Li et al. , Li, Ma et al. 2024, Ma, Li et al. 2024). This lack of granularity can obscure important intra-community differences and patterns. Also, studies frequently emphasize spatial disparities without adequately exploring temporal dynamics, such as the speed of recovery and its variability across communities (Li, Jiang et al. 2024). The interplay between spatial and temporal factors remains underexplored.

This study examines power outage impacts and recovery dynamics across Zip Codes in the Houston metropolitan area in the context of the 2024 Hurricane Beryl, a highly destructive storm that significantly affected the Houston metropolitan area. Our analysis seeks to capture various features of power outage impact and recovery, considering the duration, intensity, and spatiotemporal variation of these disruptions and the subsequent recovery speeds. Given the urban context of Houston—its diverse neighborhoods, varying infrastructure conditions, and socioeconomic profile (Ma, Li et al. , Ma, Blessing et al. 2024, Ma and Mostafavi 2024)—the study also evaluates key factors that shape the severity of the impacts and recovery timelines, including population density, median income, proximity to the hurricane path, and urban form. We are particularly interested in understanding the extent to which these factors interact to create distinct spatial patterns of impact and recovery in Houston.

Accordingly, this study seeks to answer the following research questions: (1) What is the extent of severity of power outages and their impacts on the communities during Hurricane Beryl in the Houston metropolitan area? (2) To what extent do the severity and recovery speed of power outages vary across different neighborhoods? (3) What was the extent of disparity in the severity and recovery speed of power outages across different sub-populations (e.g., income groups)? (4) To what extent do features of neighborhoods, such as development density, tree canopy, elevation, and proximity to the path of hurricane explain the variations in the severity and recovery speed across different neighborhoods? By answering these questions, this study seeks to advance our ability to quantify the severity of power outage events using multiple outages features and to further our understanding of how power outages manifest across different neighborhoods and socioeconomic groups, and how neighborhood features interact with extreme weather events to shape outage severity and recovery. This quantification of impacts and deeper insight regarding spatiotemporal variations of power outages would enable more targeted, equitable, and resilient strategies to mitigate the impacts of future power outages.

This study contributes to the body of knowledge and practices in four important ways. First, the study empirically characterizes power outages from a historic hurricane event in a large, complex metropolitan region, providing critical insight into infrastructure vulnerabilities within a high-density urban context. Second, the analysis captures the spatiotemporal dynamics of outages through a suite of metrics—ranging from outage duration to restoration rates—thereby enabling a more holistic representation of disruption severity. Third, by leveraging high-resolution outage data at both fine geographic scales (Zip Codes) and frequent time intervals, the analysis uncovers spatial disparities that would otherwise remain obscured in coarser datasets. Finally, it systematically examines the influence of socioeconomic, urban development, and environmental factors—such as income, population density, and canopy cover—in shaping the observed spatial disparities in outage impacts and recovery timelines. Combined, these contributions address the critical gaps in existing literature and lay the groundwork for more targeted, equitable, and resilient strategies to mitigate future power disruptions.

## 2. Study context and workflow

### 2.1 Study context

Hurricane Beryl impacted the Houston area in early July 2024, making landfall near Matagorda, Texas, on July 8 as a Category 1 hurricane with sustained winds of approximately 80 mph (Santini, Armas et al. 2024). The storm brought significant rainfall to the region, with some areas receiving more than a foot of rain, leading to widespread flooding. Wind gusts were notably strong, with certain locations experiencing gusts up to 107 mph, causing major damage to power poles, distribution lines, and regulator banks (Santini, Armas et al. 2024). The combination of heavy rainfall and high winds resulted in substantial power outages, affecting more than 2.7 million customers in the Houston area (Michael Zhang July 11, 2024). The storm's impact and the resulting power outages were further exacerbated by subsequent heat advisories, complicating recovery efforts and posing additional risks to the affected population (Pual Arbaje 2024). Figure 1 illustrates the path and areas directly affected by Hurricane Beryl in the Houston area, Texas. The blue-shaded regions represent the "hurricane cone," indicating the areas that are directly impacted by the hurricane. The hurricane's path is depicted by the solid dark blue line visible on the western side of the Houston area, and nearly half of the county's Zip Code Tabulation Areas (ZCTAs) fall within the directly affected zone, highlighting the widespread impact of Hurricane Beryl.

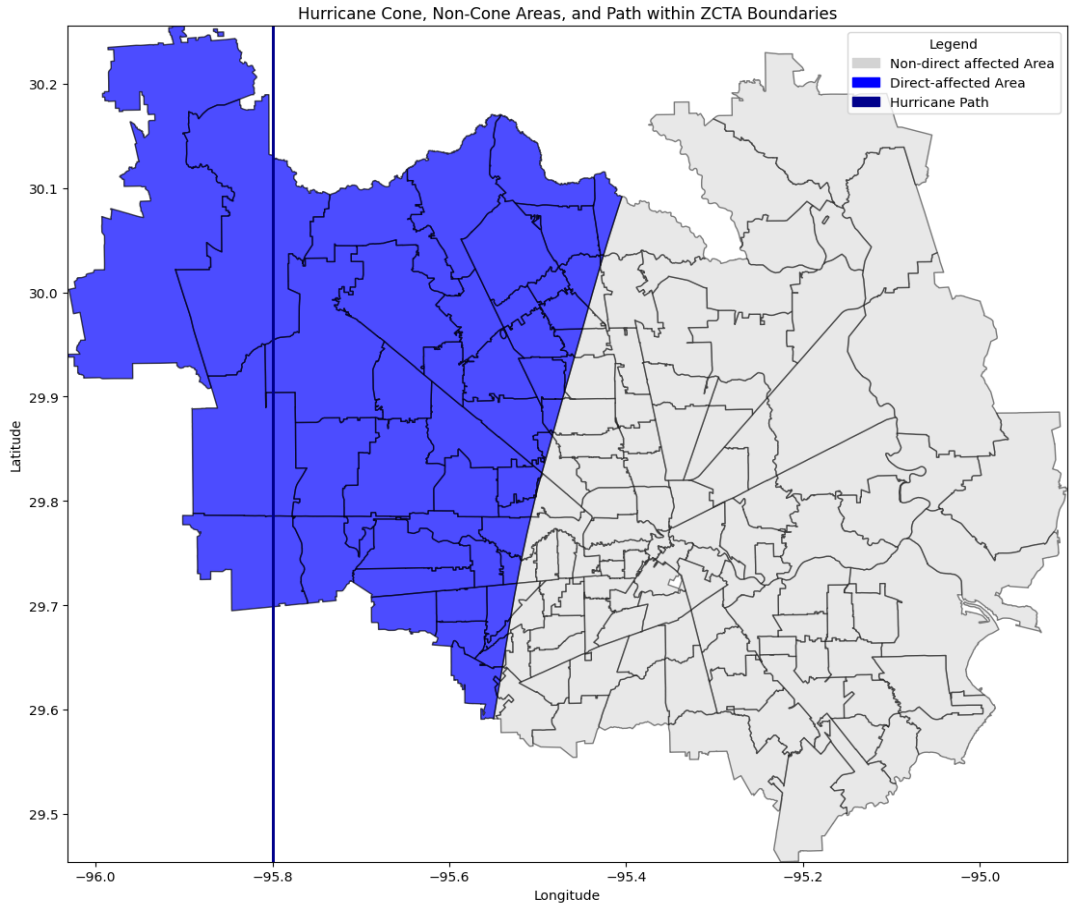

**Figure 1. Path and affected areas of Hurricane Beryl.** The solid blue line vertical is the center of the path of Hurricane Beryl. The light gray areas indicate directly affected regions, outside of the hurricane cone but within the Houston area. The solid dark blue line represents the trajectory of Hurricane Beryl as it moved through the region. (Data courtesy KMZ Viewer (KML January 13, 2025)).

**2.2 Workflow of the study**

The study followed the analyzing steps shown in Figure 2. Our analysis began with the integration of multiple datasets, including data recorded at 15-minute intervals from June to September 2024 by the Environment for Analysis of Geo-Located Energy Information (EAGLE-I™) platform (U.S. Department of Energy), a geographic information system and data visualization platform developed by Oak Ridge National Laboratory (ORNL) to provide high-resolution customer outage data. Outage hours were calculated by multiplying the number of affected customers by 0.25 hours (15 minutes), providing an estimate of customer-out hours at each 15-minute interval. We incorporated additional socioenvironmental variables, including population density, tree canopy, median and mean income, road density, elevation, distance to the hurricane path, and cone data. Our workflow is structured into three key components: dimensions development, data analysis, and statistical analysis. During dimensions development, we categorized outage impacts and recovery dimensions each into three categories. For impacts, we analyzed severity, scale, and impact duration. For recovery, we developed metrics for restoration rate, resilience, and recovery duration. In the data analysis phase, we employed K-means clustering, box plots, regression analysis, and decision-tree modeling to examine the relationships between socioenvironmental predictors and impact and recovery metrics. Finally, in the statistical analysis phase, we conducted detailed impact and recovery analyses to identify significant predictors of disaster outcomes, as well as disparity analyses to understand inequities in disaster impact and recovery across clusters.

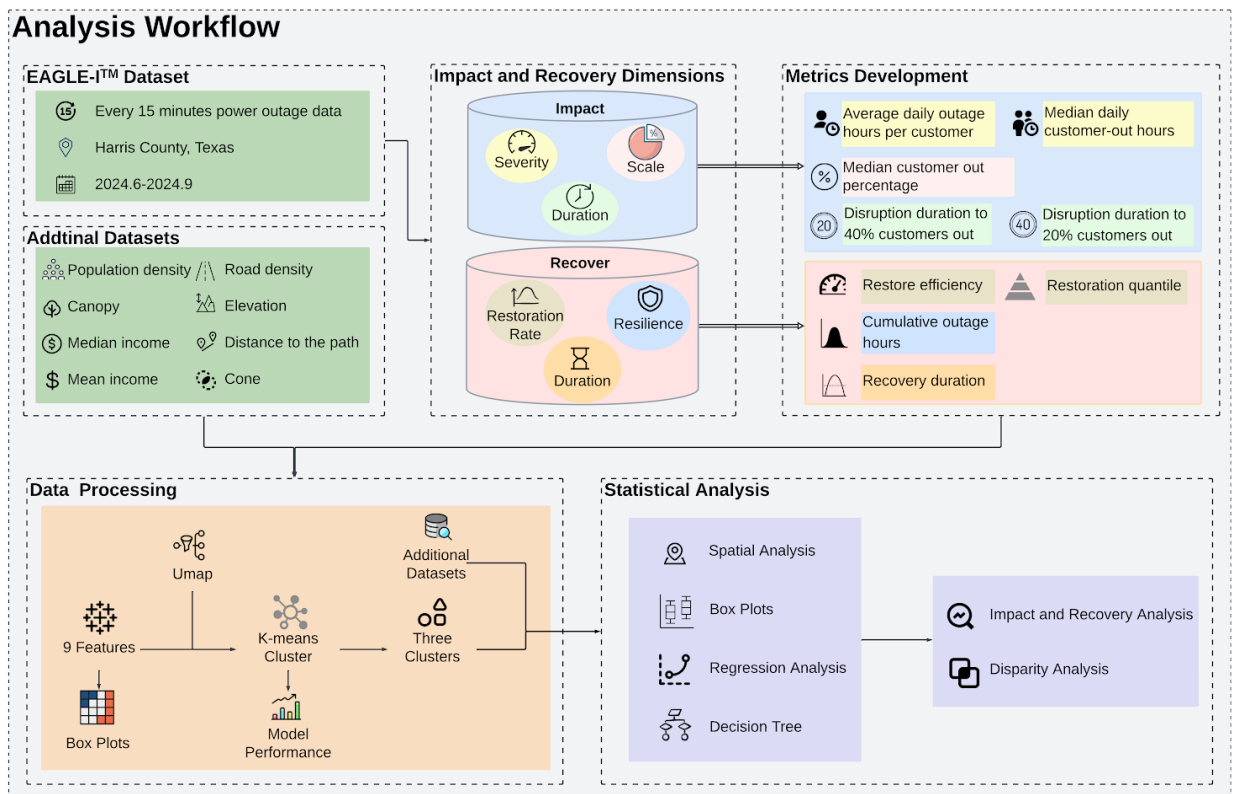

**Figure 2. Overview of the workflow and components**

### 3. Data Analysis and Results

#### 3.1 Spatiotemporal characteristics of power outages

To analyze the outage and recovery dynamics during the hurricane period, we utilized the EAGLE-I™ customer outage data aggregated by date to generate cumulative outage, cumulative restore, and resilience curves. We grouped the data by date and computed the total number of customers experiencing outages for each day, providing the foundation for tracking daily and cumulative trends. To reflect the persistent nature of outages, we calculated a modified cumulative outage curve. This curve was generated by determining, for each day, the maximum of the current day's outages and the previous day's cumulative outage value. This approach ensured that the cumulative outage accurately captured the highest sustained outage levels. Next, we calculated daily restoration values as the decrease in the number of outages from the previous day, and we constructed a cumulative restore curve by summing these daily values. To quantify the recovery progress relative to the outages experienced, we defined the resilience curve as the difference between the cumulative restore curve and the cumulative outage curve represented mathematically as

$$C(t) = R(t) - O(t) \qquad \text{Eq.1}$$

where $C(t)$ is the resilience curve, $R(t)$ is the cumulative restore, $O(t)$ is the cumulative outage, and $t$ represents days.

We then defined the final date as the date when the restore catches the cumulative outage. The outage rate was calculated as the maximum cumulative customer-out divided by the number of days between the start of the outage and the final date. The restore rate was determined as the maximum cumulative customer-out of restore divided by the duration from the first restoration date to the final date. We defined the restore efficiency as the outage rate over the restoration rate.

From Figure 3, we can see that the outage curve (blue) shows a sharp and rapid increase initially, indicating significant and sudden outages. This curve plateaus at approximately $1.3 \times 10^8$ cumulative customers, suggesting the peak outage was reached quickly and did not escalate further beyond this point. In contrast, the restore curve (red) progresses more steadily, indicating a gradual restoration process. The restore curve converges with the outage curve at the end, reaching approximately $1.3 \times 10^8$ cumulative restored cumulative customers, verifying that full restoration was achieved. The resilience curve (green) begins with negative values as the outages initially outweigh the restorations. We also calculate the area between the outage and restoration curve, which is also equal to the resilience curve to x-axis. This area is the cumulative customer-out, representing the net cumulative customer-out over time. The final date when the restoration catches the cumulative outage is July 24, 2024. We also plotted the al distribution of outage rate and restoration rate (Figure 4).

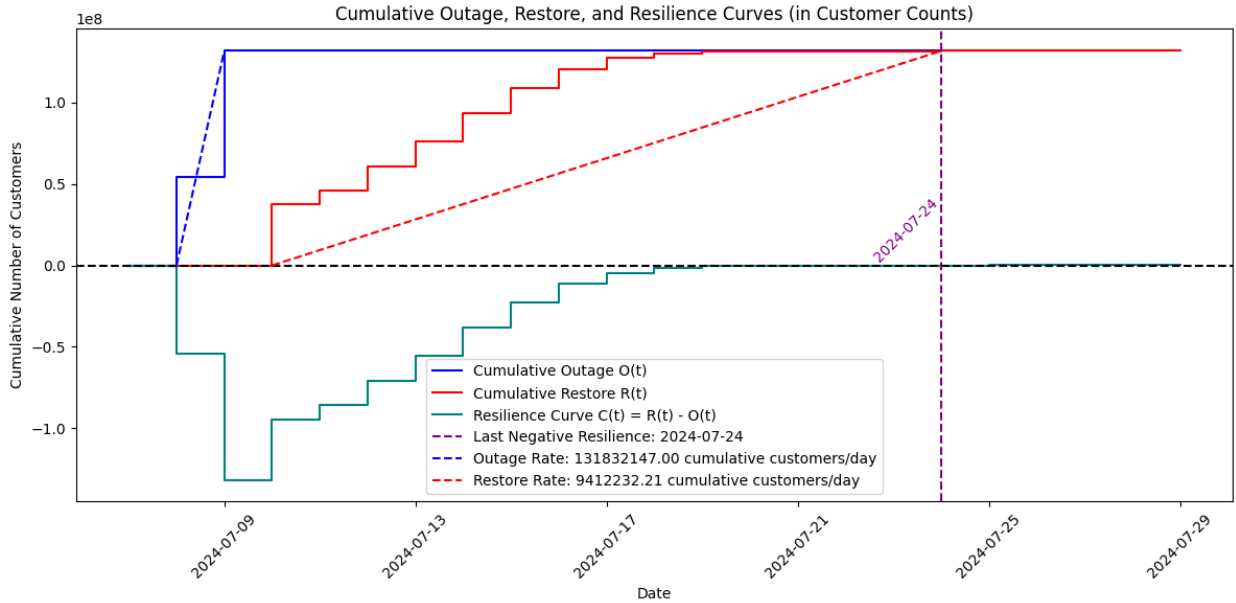

**Figure 3. Overview of the power outage and recovery dynamics.** The cumulative outage curve (blue) depicts the total number of cumulative customers affected by outages over time, while the restore curve (red) shows the total cumulative number of customers restored within the same timeframe. The resilience curve (green) highlights net recovery progress, providing a visual representation of the restoration trajectory. Dashed lines indicate calculated outage and restore rates, serving as quantitative benchmarks to assess recovery performance. The vertical purple line marks the point when restoration efforts fully match the total outages, signaling the completion of recovery.

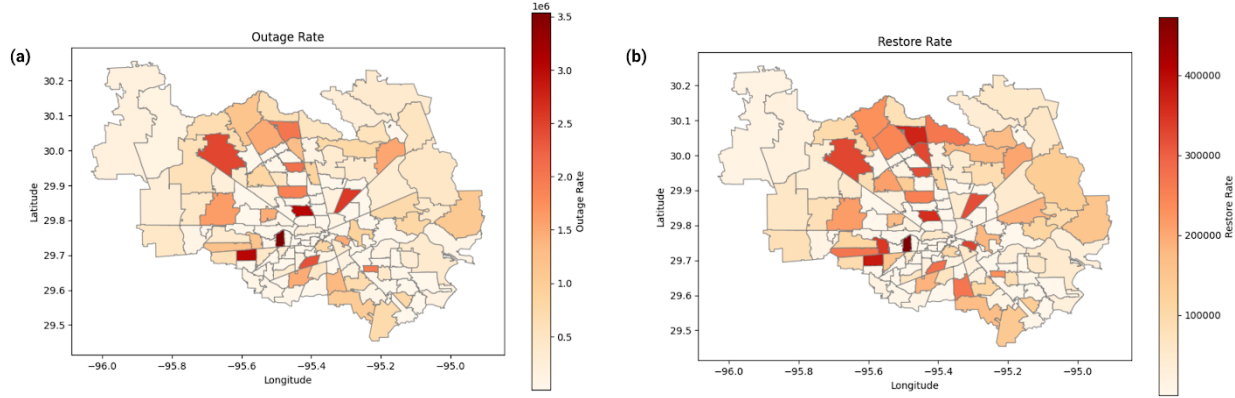

**Figure 4. The outage rate and restore rate in the Houston area, Texas, across ZCTAs.** The left plot illustrates the spatial distribution of the outage rate, calculated as the cumulative number of customers affected per day during the study period. Darker red areas represent regions with higher outage rates, indicating significant disruptions to power service. The right plot depicts the restore rate, defined as the cumulative number of customers whose power was restored per day

in each ZCTA. The darker red areas correspond to higher restore rates, highlighting regions where recovery efforts were swift and effective.

Figure 5 plots the trends of daily total customer-out hours from July 1, 2024, through September 1, 2024, with a red dashed baseline representing pre-event conditions. The baseline was calculated as the mean daily total customer-out hours between June 1, 2024, and July 1, 2024, capturing the normal, pre-disruption state. Red points mean the daily total custom-out hours is equal to or less than the pre-disruption state. On July 8, the total outage hours showed a sharp increase, coinciding with the impact of Hurricane Beryl. The outage hours peaked shortly after, reflecting the widespread power disruptions caused by the event. Following this peak, a steady decline in outage hours starting from July 10 is observed as recovery efforts progressed. The first instance where the daily outage hours returned to the baseline level occurred on July 27, 2024, indicating an overall recovery to pre-event conditions. The date is three days longer than the restoration date, which is July 24. Subsequent data points show continued stabilization, with some days remaining at or below the baseline, signifying that the recovery phase was nearing completion by the end of the observation period.

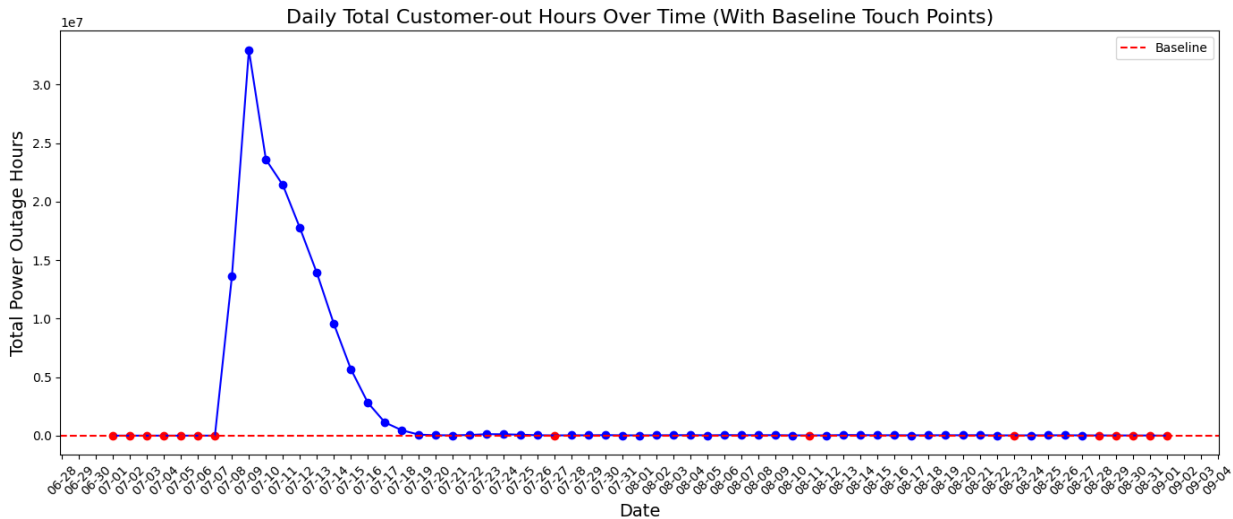

**Figure 5. Daily Total Customer-Out Hours Over Time.** The continuous line illustrates the temporal trend of daily total customer-out hours, capturing fluctuations throughout the period. A dashed red line represents the baseline, serving as a reference for pre-event levels. Points at or below the baseline are highlighted in red, indicating recovery to normal levels, while points above the baseline are marked in blue, signifying periods of elevated outage levels.

Figure 6 shows the spatial distribution of total customer-out hours across the Houston area during Hurricane Beryl. We aggregated the data by summing the "outage hours." which is also the area under the curve in Figure 5 for each ZCTA area. Between July 8, 2024, and July 27, 2024, the total customer-outage hours amounted to 143,455,003.25 hours.

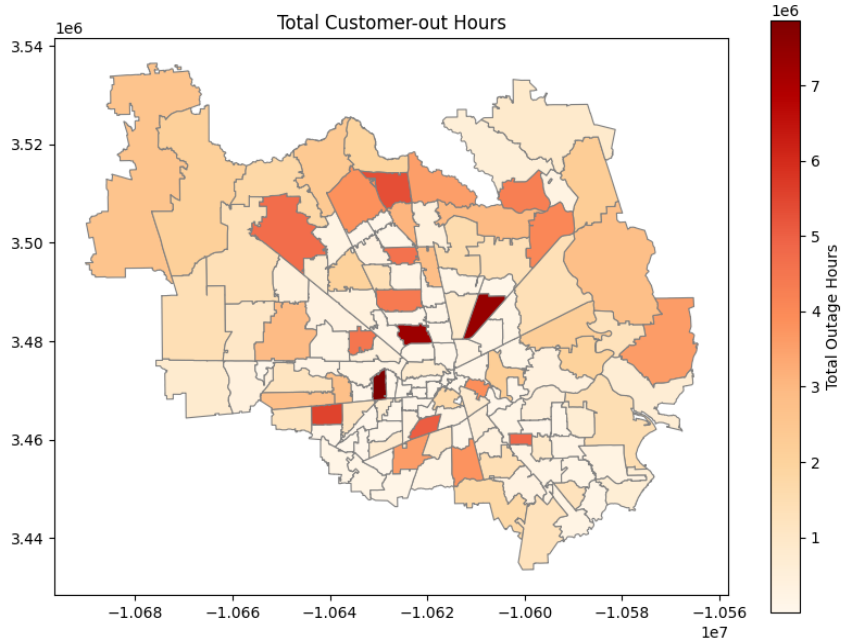

**Figure 6. Spatial Distribution of Total Customer-Out Hours.** The map displays the distribution of total customer-out hours across the ZCTA areas ranging from approximately 1 million hours (lightest shades) to more than 7 million hours (darkest shades). Areas with the highest total customer-out hours, represented in dark red, are concentrated in the central and northern parts of the region. In contrast, the peripheral regions, particularly in the eastern and western areas, show the lightest shades, indicating fewer customer-out hours due to shorter outage durations or lower population densities.

### 3.2 Power outage impact and recovery features

We developed nine features from the outage data and divided them into two key dimensions: impact and recovery (Figure 7). The impact dimension is developed from three categories: (1) severity: average daily outage hours per customer, median daily customer-out hours; (2) scale: median customer-out percentage; (3) duration: the length of time between the start of power outage to the point at which 40% of customers are without power, and to the point at which 40% of customer's power has been restored. The Recovery dimension can also be divided into three categories: (1) restoration rate: restoration quantile; (2) restore efficiency; (3) resilience: cumulative customer-out; duration: recovery duration.

**Table 1. Illustration of the power outage impact and recovery features**

| Feature Category | Feature Name | Feature Description | Feature Interpretation |
|---|---|---|---|

| Impact | Average daily outage hours per customer | (Daily total customer-outage hours / daily maximum number of customers out per ZCTA)/ disruption days per ZCTA | Reflects the severity of the disruption for each customer. |
|---|---|---|---|
| | Median daily customer-out hours | Daily total customer-out hours / disruption days | Captures the daily severity of customer-out hours |
| | Median customer out percentage | Median of (daily customers out / maximum customers out) over disruption days | Indicates the scale of the disruption, showing how widespread the impact is. |
| | Disruption duration to 40% customers out | Number of days from July 8 to when 40% of customers still power outage | Reflects the timeline of disruption and the subsequent impact duration. |
| | Disruption duration to 20% customers out | Number of days from July 8 to when 20% of customers still has power outage | Reflects the timeline of disruption and the subsequent impact duration. |
| Recovery | Restoration quantile | Quantile of restoration rates based on the percentage change in cumulative customer-out hours over a rolling seven-day period | Reflects the peak efficiency of restoration efforts. |
| | Restore efficiency | Restore rate/outage rate | Indicates the balance between recovery speed and outage speed. |
| | Net cumulative customer restoration | Area between the outage and restoration curve, representing the net cumulative customer-out over time per Zip Code | Quantifies the gap between outages and recovery, reflecting system resilience. |
| | Recovery duration | The number of days between the first exceedance of total customer-out hours baseline and the first return to or below the baseline | Reflects the temporal extent of the impact, capturing how long the disruption lasts. |

### 3.2.1 Power outage impact features

#### 3.2.1.1 Average daily outage hours per customer

To calculate the average daily outage hours per customer for each Zip Code (Figure 8(a)) during the disruption period (July 8 through July 27, 2024), we first aggregated the total customer-out hours, which represents the cumulative duration of outages experienced by all customers in each area. Then, to estimate the number of customers, we identified the maximum number of customers-out in each Zip Code for each day, capturing the highest count of affected individuals during daily peak outages. Using these metrics, we calculated the average daily outage hours per customer for each Zip Code by dividing the daily total customer-out hours by the daily maximum number of customers out. Finally, to represent the overall impact during the event, we computed the average values of the feature over disruption days for each Zip Code. The overall average daily outage per customer was 12.41 hours, resulting in a total average outage per customer of 248.2 hours over 20 days. Figure 8(a) displays the spatial distribution of the average daily outage hours per customer. We observed that the average daily outage hours ranged from approximately 9 to 15 hours. High average outage hours (red) are concentrated primarily in the central and southwestern areas of the region. Specifically, some Zip Codes in the central areas exhibit values exceeding 14 hours, indicating prolonged disruptions for customers in these areas. Conversely, the lighter shades, representing lower average outage hours (closer to 9 to 11 hours), are predominantly distributed along the periphery, particularly in the northern and eastern parts of the region.

### 3.2.1.2 Median daily customer-out hours

Median daily customer-out hours represent the median value of daily cumulative outage hours (Figure 8(b)). First, we calculated the daily total outage hours by summing the outage hours, which represents the cumulative outage duration across all affected customers. Next, we calculated the median daily outage hours for each Zip Code over the selected period (July 8 through July 27). Figure 8(b) illustrates the median daily customer-out hours across the Houston area during the disruption period, with a gradient of red shades representing varying outage durations. Darker shades indicate areas with higher median outage hours, while lighter shades represent less severe impacts. The most affected regions are predominantly concentrated in the northeastern part of the Houston area, where the median daily customer-out hours exceeded 60,000 hours, indicating prolonged outages in these areas. In addition, a smaller cluster of high values is observed in the southern region.

### 3.2.1.3 Median customer-out percentage

For each Zip Code, the maximum number of customers affected by outages during the disruption period was identified and recorded in the raw dataset. The customer outage percentage (Figure 8(c)) was calculated for each day as the ratio of customer-out to customer-max. This percentage represents the proportion of customers in each Zip Code which experienced outages on any specific day and time. To summarize the impact over time, the median customer out percentage for each Zip Code was computed during the period July 8 through July 27. The maximum daily median percentage of customers out of power was 84.25%. Figure 8(c) displays the spatial map of the feature, with a red gradient indicating the severity of outages. Darker shades represent Zip Codes with higher median outage percentages, while lighter shades indicate less severe impacts. The central part of the county experienced the highest median outage percentages, with some areas reaching as high as 60%, indicating that a significant portion of customers in these regions was affected during the outage period. Additionally, elevated outage percentages are observed in parts of the southeastern and northeastern regions, reflecting pockets of concentrated disruption. In contrast, the western regions of the county experienced lower outage percentages, as indicated by the lighter red shades.

**3.2.1.4 Disruption duration to 40% customers out and disruption duration to 20% customers out**

To calculate the days required to reach specific outage percentages (40% and 20% of customers out, Figures 8(d) and 8(e)), we focused on the maximum daily customer outage percentages for each Zip Code during the disruption period. First, the percentage of customers out was calculated daily as the ratio of customers experiencing outages to the maximum number of customers affected. The data was then grouped by Zip Code and date to determine the maximum outage percentage for each Zip Code on each day. For each Zip Code, thresholds of 40% and 20% customer outages were defined. To determine the days required to reach these thresholds, we compared the daily outage percentages against the thresholds to identify the day when the percentage was closest to the defined threshold. The time to reach these thresholds was calculated as the difference in days from the baseline date (July 8, 2024), which marks the beginning of the disruption period. Figures 8(d) and 8(e) illustrate the number of days required for each Zip Code to reduce outages to 40% and 20% of the customer-out percentage, respectively. The color gradient, ranging from light pink to dark red, represents the time to reach this threshold, with darker shades indicating longer durations. The northeastern region of the Houston area shows the most significant delays, with some areas requiring up to 14 days to reach the 40% threshold, and 16 days to reach 20%. In contrast, central and southwestern regions display shorter recovery times, typically under 6 to8 days to 40% and 20%, as indicated by the lighter shades. While the northeastern region again exhibits the most extended recovery times, the southeastern and southern parts of Houston area also show delays, with several Zip Codes taking between 8 and 12 days to reach these thresholds.

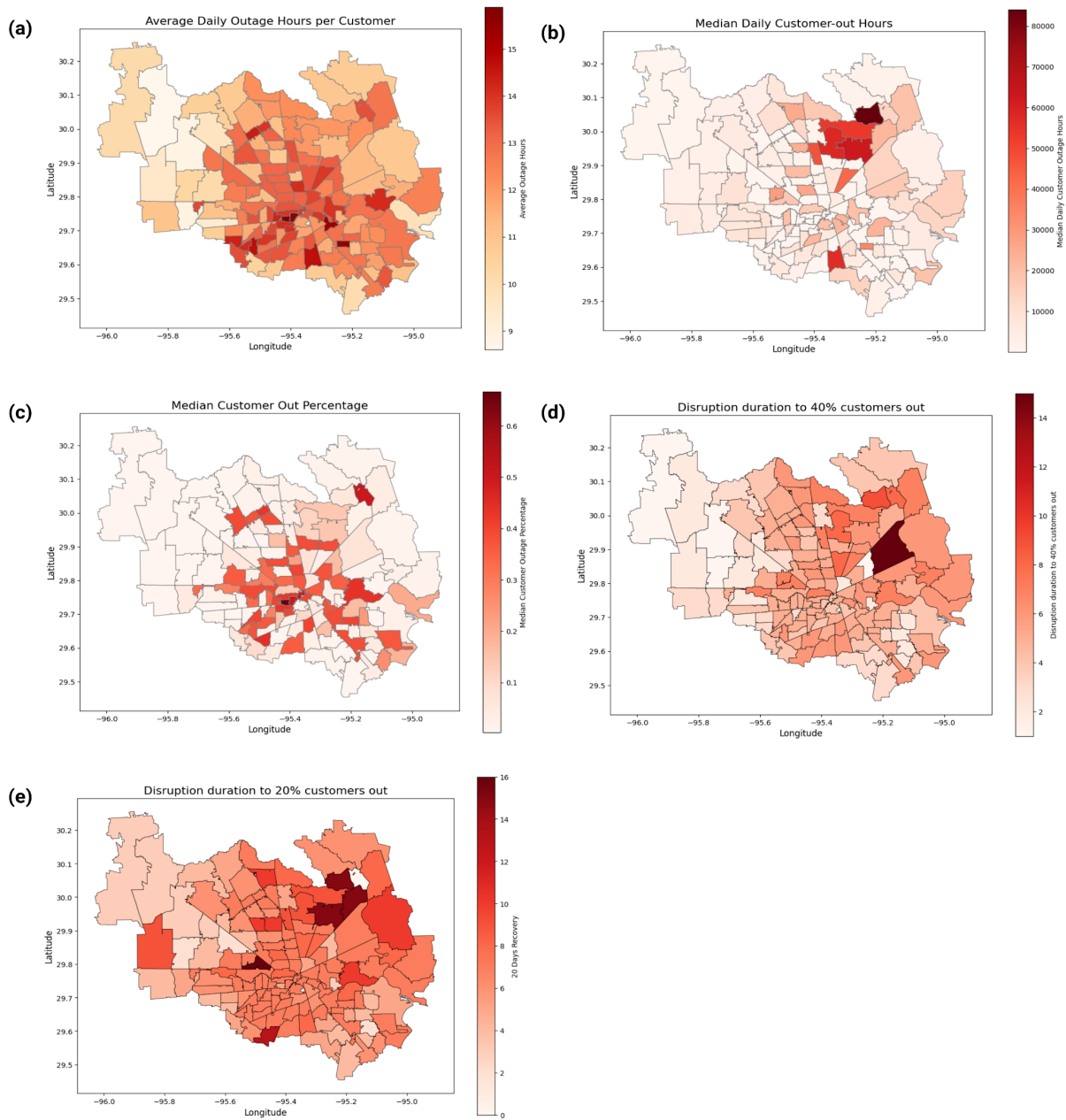

**Figure 8. Spatial distribution of the power outage impact features. (a) Average outage hours per customer; (b) Median daily customer-out hours; (c) Median customer out percentage; (d) Disruption duration to 40% customers out; (e) Disruption duration to 20% customers out.** Darker red areas indicate Zip Codes with higher values of the features, signaling areas where outages were more prolonged or severe. Conversely, lighter regions represent areas with less severe disruption.

### 3.2.2 Power outage recovery features

#### 3.2.2.1 Restoration quantile

To calculate the restoration quantile, we first calculated restoration rates based on the percentage change in cumulative customer-out hours over a rolling seven-day period (Eqs. 2 and 3). Next, we divided the calculated restoration rates into four quantiles (Q1 to Q4), with Q1 representing the fastest restoration rates and Q4 indicating the slowest.

$$S_t = \sum_{i=t-6}^{t} H_i \qquad \text{Eq. 2}$$

where $S_t$ represents seven-day rolling sum of customer-out hours on day $t$, and $H_i$ represents daily customer-out hours on day $i$ (from t-6 to t).

$$R_t = \frac{S_t - S_{t-1}}{S_{t-1}} \times 100\% \qquad \text{Eq.3}$$

where $R_t$ represents restoration rate (percentage change) on day $t$, $S_t$ represents seven-day rolling sum of outage hours on day $t$, and $S_{t-1}$ represents a seven-day rolling sum of outage hours on day $t$-1.

The restoration rate quantile map (Figure 9(a)) displays the spatial distribution of restoration rates across the Houston area: blue (Q1, fastest restoration), green (Q2), orange (Q3), and red (Q4, slowest restoration). Areas in the central-eastern and peripheral parts of the county generally fell into Q1 and Q2, indicating faster restoration rates. In contrast, several Zip Codes in the northeastern and central regions are categorized as Q3 and Q4, reflecting slower recovery rates.

#### 3.2.2.2 Restore efficiency

To compute the restore efficiency (restore rate/outage rate), we analyzed customer outage and restoration data over the period from July 8 through July 30, 2024, focusing on the interplay between outage severity and recovery efforts for each Zip Code. The restore efficiency was derived as the ratio of the restore rate to the outage rate, as previously mentioned. This metric quantifies the balance between the speed of recovery efforts and the severity of the outage, with higher values indicating faster recovery relative to the outage impact. Figure 9(b) displays the restore efficiency for each Zip Code in the Houston area. Darker red represents higher restore efficiency, signifying areas where restoration efforts outpaced outage severity. In contrast, lighter shades indicate lower restore efficiency, reflecting slower recovery relative to the magnitude of the outages. From the map, we observe that the northeastern region exhibits the highest restore efficiency, suggesting efficient recovery efforts despite significant outages.

Conversely, several Zip Codes in the western and southeastern regions display lower restore efficiency, indicating slower recovery relative to outage severity.

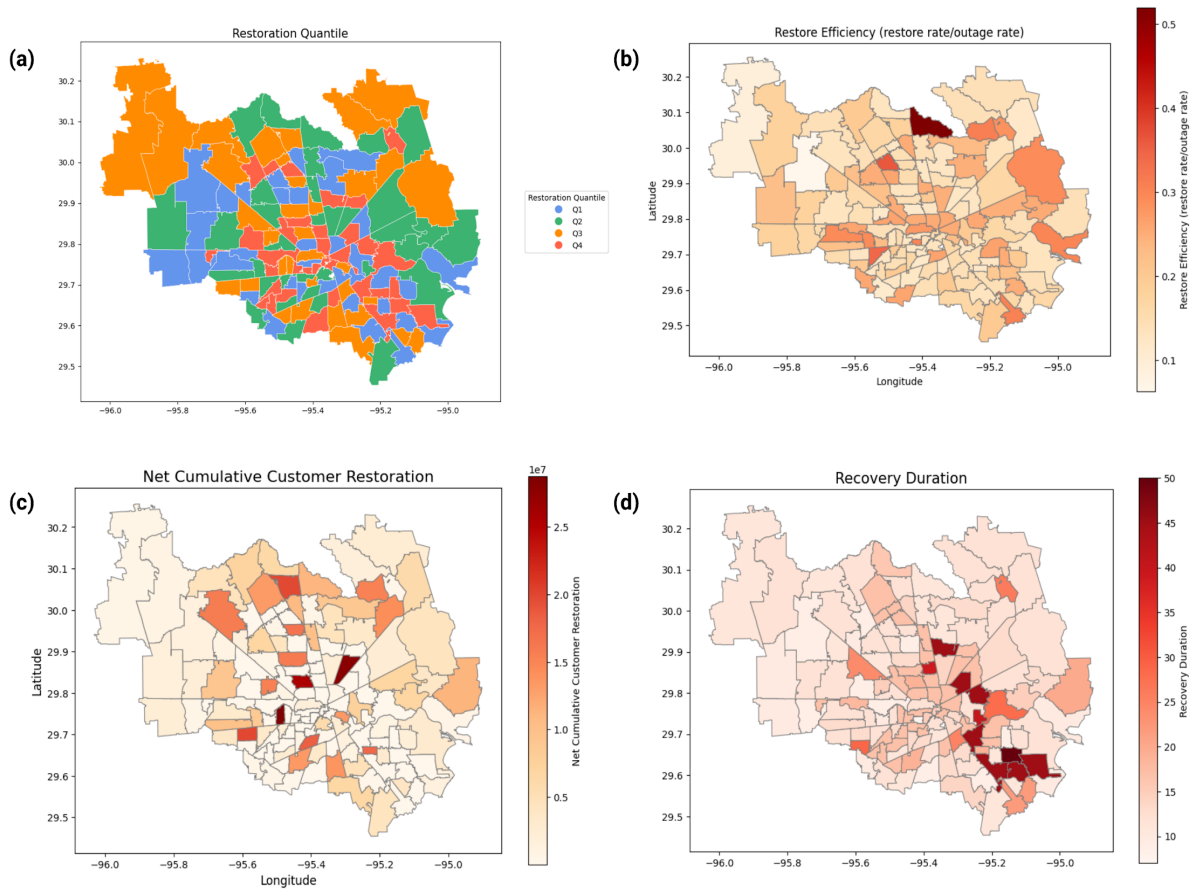

**Figure 9. Spatial distribution of the power outage recovery features. (a) Restoration quantile; (b) Restore efficiency; (c) Net cumulative customer restoration; (d) Recovery duration** Darker red areas indicate Zip Codes with higher values of the features, signaling areas where recovery was slower. Conversely, lighter regions represent areas with faster recovery.

**(c) Net Cumulative customer restoration**

The net cumulative customer restoration was determined by calculating the area between the outage and restoration curve, which is also the area of the resilience curve. Higher cumulative customer-out restoration indicates slower recovery or less effective restoration efforts relative to the extent of the outage. The cumulative customer-out map (Figure 9 (c)) provides a spatial overview of the feature; the darker red indicates higher cumulative customer-out, corresponding to larger gaps or delays in recovery. The central and southeastern regions exhibit the highest cumulative customer-out. In contrast, Zip Codes in the western regions display smaller cumulative customer-out.

**(d) Recovery duration**

We calculated recovery duration to measure the time required for fully recovery to the pre-event status. The first step involved establishing a baseline outage level for each Zip Code. This baseline was defined as the average daily outage hours during a pre-event period from June 1 to July 1, 2024, representing normal operating conditions before the disruption. By setting this baseline, we could identify deviations in outage patterns caused by the event for each area. We examined the daily outage hours for each Zip Code for the disruption period (July 8 through September 1, 2024). The outage start date was identified as the first day when outage hours exceeded the baseline; the recovery date was determined as the first subsequent day when outage hours returned to or fell below the baseline. The recovery duration was then calculated as the number of days between the outage start date and the recovery date. Figure 9(d) illustrates the spatial distribution of recovery duration across the study area. The lighter shades indicate shorter recovery durations where power restoration was achieved more quickly. These efficiently restored areas are predominantly located in the northeastern and northwestern parts of the study area. In contrast, darker red shades represent longer recovery durations, highlighting areas where power outages persisted for extended periods. These prolonged outage regions are concentrated in the central and southern portions, where recovery durations exceed 40 days.

A bar chart (Figure 10) illustrates the distribution of recovery completion dates for each Zip Code in 2024. The x-axis displays the recovery dates from July through August, while the y-axis shows the number of Zip Codes that achieved recovery on each date. We can see from the blue bars, the majority of recoveries occurred between July 19 and July 25, 2024, with peak activity observed on July 20 (32 recoveries) and July 25 (33 recoveries). Recovery activity decreased substantially after July 25, and it continued to August 27.

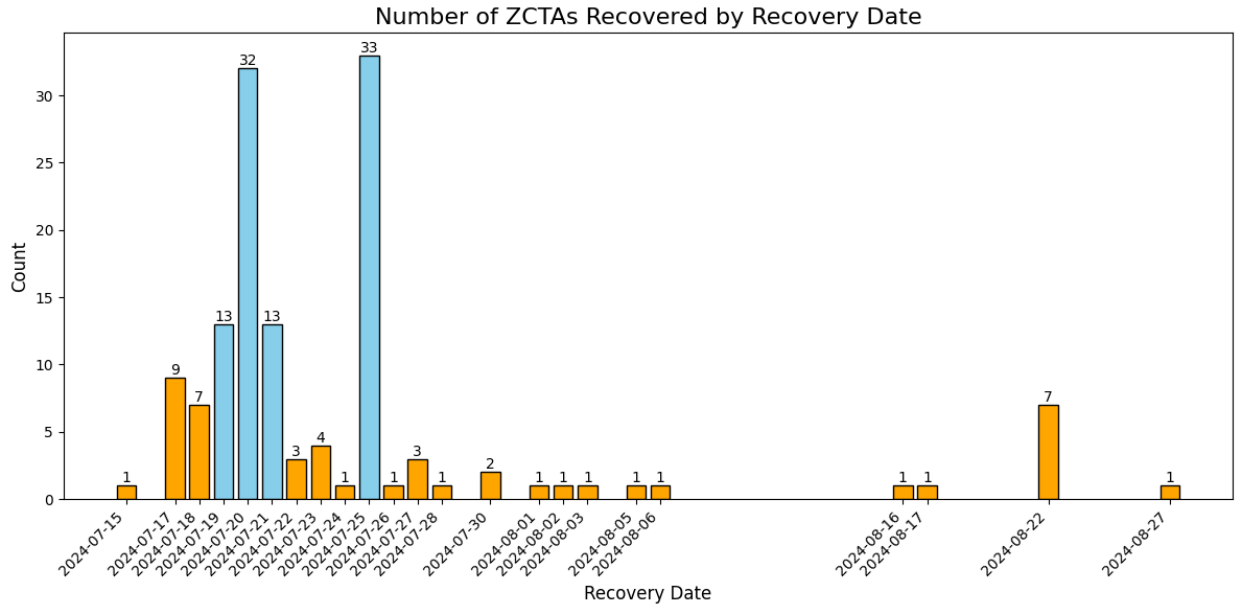

**Figure 10. Number of ZCTAs recovered by recovery date.** Each bar corresponds to a specific date. The height of the bar reflects the number of recoveries on that date. The bars are color-coded to differentiate recovery levels: blue bars represent dates with more than ten recoveries, indicating significant recovery activity, while orange bars denote dates with ten or fewer recoveries.

### 3.3. Statistical analysis of spatial disparities

To analyze the spatial disparities in power outage impacts and recovery, we adopted K-means clustering to group Zip Codes based on the features. Before conducting the clustering analysis, we performed a Pearson correlation analysis to examine the relationships between features and ensure minimal multicollinearity among them. Correlation coefficients range from -1 to 1, with values approaching 1 indicating strong positive correlations, values near -1 indicating strong negative correlations, and values close to 0 indicating weak or negligible correlations (Sedgwick 2012). The results (Figure 11) show that all correlation coefficients remained below 0.7, indicating suitable conditions for performing the clustering analysis. Notable results from the correlation matrix include a strong positive correlation (r = 0.6) between median daily customer-out hours and net cumulative customer restoration, suggesting that geographic areas experiencing higher daily outages typically correspond to larger resilience areas. In contrast, features such as median customer-out percentage and net cumulative customer restoration show a strong negative correlation (r $\approx$ -0.5), suggesting that regions with a higher percentage of outages are likely to recover faster.

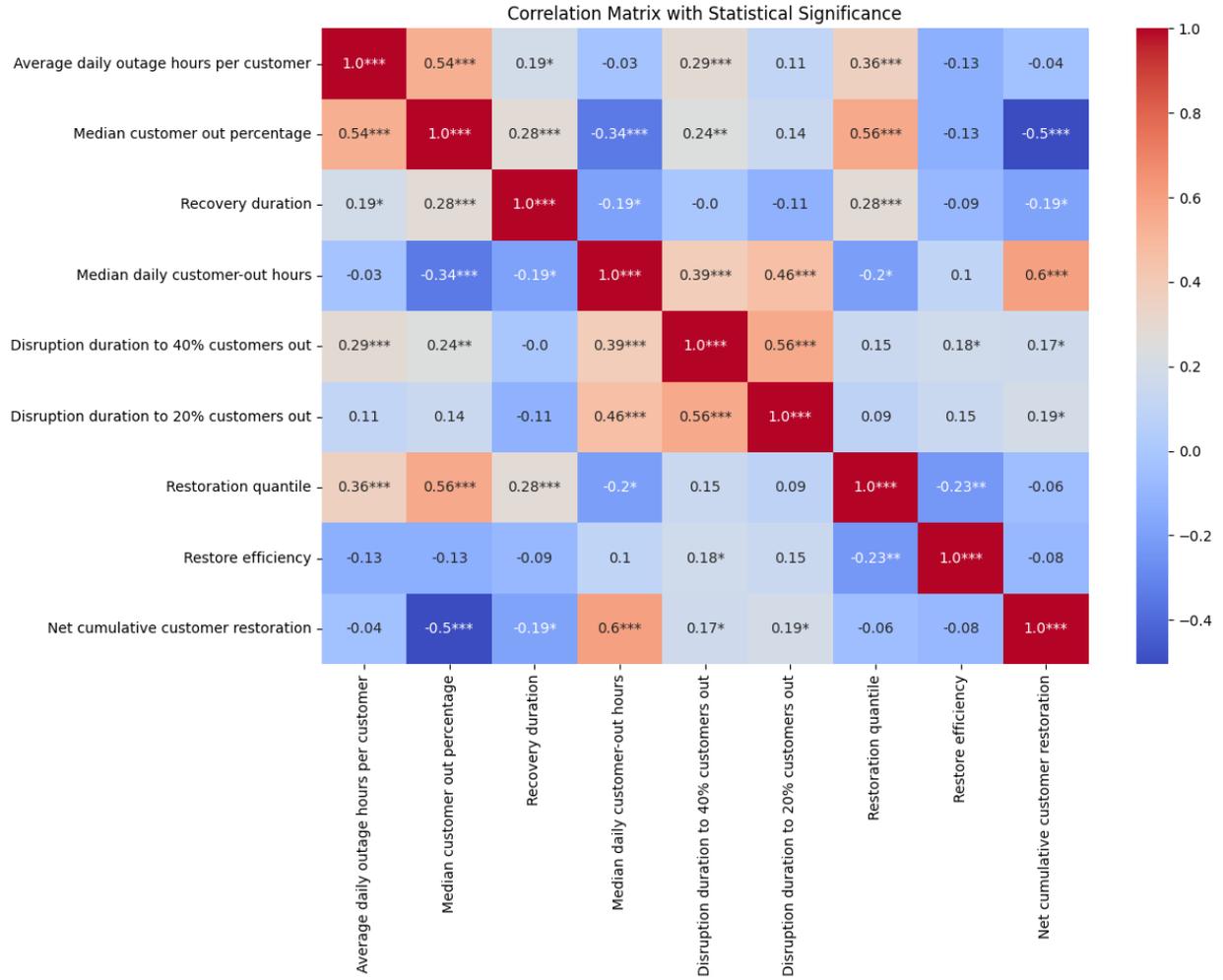

**Figure 11. Pearson correlation matrix between power outage and recovery features. The number within the cell indicates the correlation coefficient. ***$p<0.001$, **$p<0.01$, *$p<0.05$.**

We then applied uniform manifold approximation and projection (UMAP) to reduce the high-dimensional feature space into two dimensions while preserving the underlying structure and relationships within the data (McInnes, Healy et al. 2018). The K-means algorithm was subsequently applied to the UMAP-transformed data (Eq. 4). K-means iteratively assigned each data point to the nearest cluster centroid while minimizing the variance within clusters and maximizing the variance between clusters (Hamerly and Elkan 2003). The final cluster assignments were added as a categorical variable, enabling us to incorporate the clustering results into subsequent analyses.

$$J = \sum_{i=1}^{K} \sum_{\chi \epsilon C_i} ||x - \mu_i||^2 \qquad \text{Eq.4}$$

where K represents the number of clusters, $C_i$ is the set of points in cluster, x is an individual data point, $\mu_i$ denotes the centroid of cluster, $\|x - \mu_i\|^2$ is the squared Euclidean distance between a data point and its cluster centroid.

Three clusters were finally generated with a silhouette score of approximately 0.574. To validate the robustness of the clustering, we conducted ANOVA (analysis of variance) tests across the three clusters. ANOVA results demonstrated statistically significant differences (p-value < 0.05) across clusters, which confirmed that the clusters captured meaningful and distinct groupings in the data. Figure 12 illustrates the spatial distribution of three distinct clusters identified through our analysis, with each Zip Code color-coded according to its cluster assignment. We also created boxplots in Figure 13 to compare key impact and recovery features across the three identified clusters. Each cluster exhibits distinct characteristics that provide insights into the varying patterns of power outage impact and recovery processes.

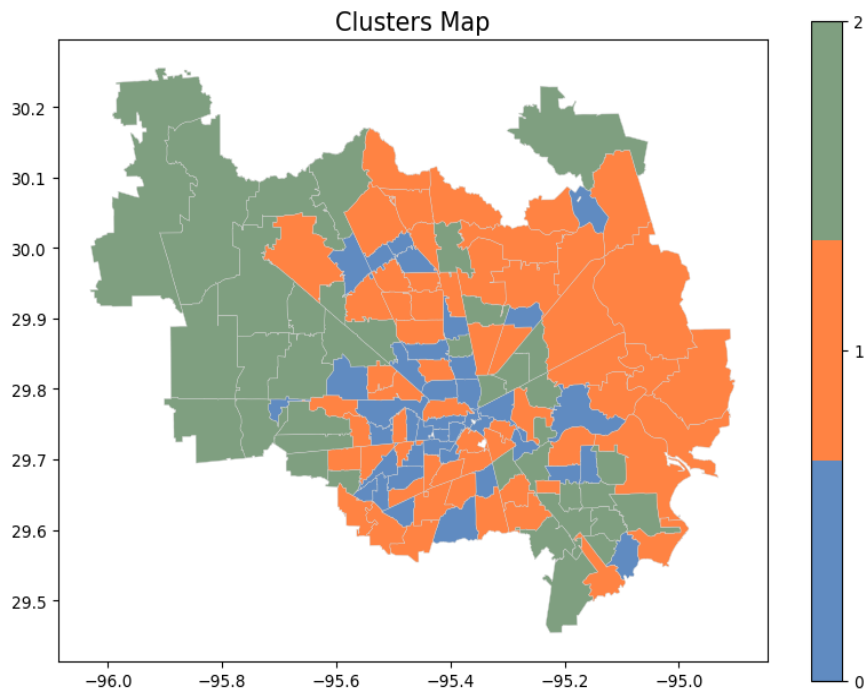

**Figure 12. Spatial distribution of the three clusters regarding power outage impact and recovery in the Houston area.** We applied the K-means to implement the clustering analysis, and the nine features were reduced to two dimensions using UMAP before clustering. The three clusters are represented by different colors.

**Cluster 0: High-impact outages with slower recovery**

Cluster 0 (blue) is predominantly concentrated in the central and southern portions of the study area. Cluster 0 demonstrates the most severe outage impacts in the study area. This cluster records the highest average daily outage hours per customer, approaching 14 hours, and exhibits the greatest median percentage of affected customers. In addition, the delayed disruption period as evidenced by extended time frames for reducing outages to both 40% and 20% of initially affected customers. While the median daily customer-out hours are lower than those in other clusters, suggesting a distinct temporal distribution of service disruptions, the recovery process shows moderate delays. These delays are evidenced by elevated restoration quantiles over seven-day moving averages and lower restore-to-outage restore efficiency compared to other clusters. Despite having a relatively small resilience area, indicating better alignment between outage and restoration patterns during the event, the recovery duration exceeds that of cluster 1 while remaining shorter than cluster 2. These characteristics point to substantial vulnerabilities in both infrastructure and response capabilities, emphasizing the need for targeted improvements in system resilience and restoration processes.

**Cluster 1: Moderate impact with efficient recovery**

Cluster 1 (orange), occupying primarily the eastern section of the Houston area, exhibits moderate outage impacts coupled with superior restoration efficiency. The cluster shows daily outage hours per customer average of approximately 12 hours, and the highest median daily customer-out hours, indicating widespread concurrent service disruptions. The proportion of affected customers falls between the higher level in cluster 0 and the lower level in cluster 2. Cluster 1 has moderate timeframes for reducing outages to key thresholds. Recovery features demonstrate strong performance: lower restoration quantiles indicating faster average service restoration and higher restore-to-outage restore efficiency, showing more effective repair operations than cluster 0. Although this cluster has a larger resilience area, suggesting greater separation between outage and restoration curves, it achieves the shortest overall outage duration among all clusters. These results indicate that despite moderate initial disruptions, effective coordination and resource allocation enable rapid system-wide recovery.

**Cluster 2: Low initial impact but prolonged restoration**

Cluster 2 (green) is most prevalent in the southeastern and western regions of the study area. Cluster 2 shows the lowest initial impact severity but faces challenges in achieving complete recovery. Average daily outage hours per customer remain below 12 hours, accompanied by minimal daily customer-out hours and the lowest percentage of affected customers. The cluster achieves shorter timeframes for reducing outages to both 40% and 20% thresholds, indicating an effective initial response. However, these early advantages do not lead to swift overall recovery. Despite maintaining steady daily restoration rates, as evidenced by a balanced restoration efficiency and the lowest restoration quantile, this cluster records the longest total recovery duration. Its moderate resilience area, falling between clusters 0 and 1, indicates typical separation between cumulative outage and restoration patterns. These findings suggest that while initial storm resistance is strong, underlying systemic constraints may impede complete service restoration.

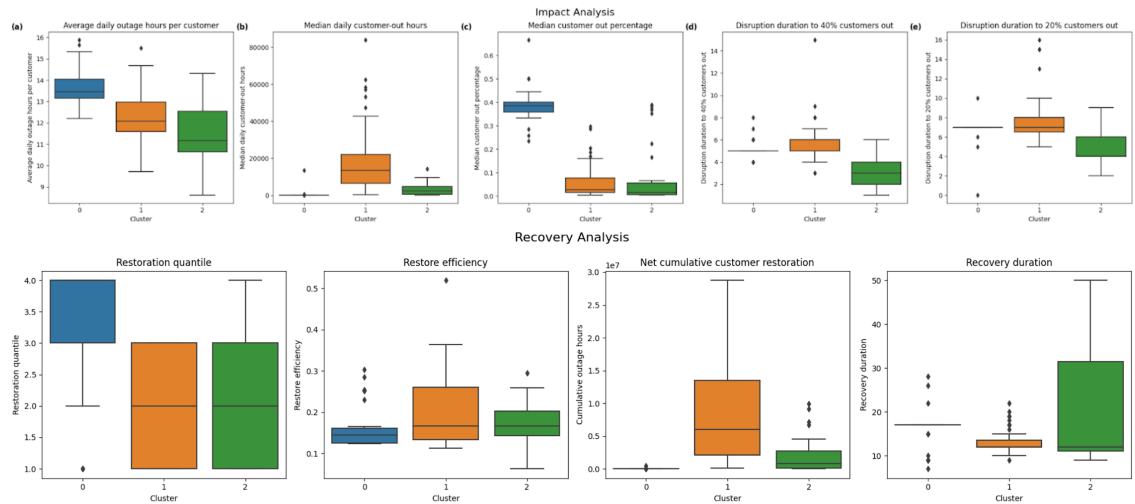

**Figure 13. Box plots of the outage impact and recovery features Values across Clusters. (a) Average daily outage hours per customer; (b) Median daily customer-out hours; (c) Median customer out percentage; (d) Disruption duration to 40% customers-out; (e) Disruption duration to 20% customers-out. (d) Restoration quantile; (e) Restore efficiency; (f) Net cumulative customer restoration; (g) Recovery duration.**

In the next step, we created the spatial distribution of quantitative differences across clusters for the impact and recovery features in (Figures 14 and 15). For average daily outage hours per customer, cluster 0 exhibited the highest values, reflecting prolonged outages in the areas. Cluster 1 showed intermediate values, while cluster 2 displayed the lowest values. For median daily customer-out hours, cluster 1, especially the northern part in this cluster, exhibited significantly higher values, whereas clusters 0 and 2 showed comparatively lower values. The median customer outage percentage highlights differences in the extent of customer impact. Cluster 0 demonstrated the highest values of the feature, especially in the central region, indicating widespread impacts within these areas. In contrast, clusters 1 and 2 showed the lowest values, suggesting that outages in these areas do not affect a large portion of customers. The disruption days to 40% customer-out and days to 20% customer-out are similarly distributed in the map and they reveal significant disparities across clusters in recovery timelines. Cluster 1, mainly in the northeastern regions, consistently showed longer durations, suggesting delays in impact efforts. Cluster 0 demonstrated the shortest restoration times, reflecting efficient recovery processes. Cluster 2 showed moderate restoration times.

The restoration quantile reveals stark differences in recovery speed across clusters. Cluster 0, especially in central and eastern regions, consistently exhibited the highest quantile values in central regions, indicating the lowest recovery rate. Conversely, clusters 1 and 2 demonstrated lower quantile values in western and eastern regions, reflecting a quicker recovery rate. For the restore efficiency, Clusters 0 and 2, concentrated in the western regions, displayed higher recovery rates, while Clusters 1 showed slower recovery rates. The net cumulative customer restoration metric highlights cumulative delays in recovery. Cluster 1 exhibited the largest net cumulative customer restorations, located in the northern and eastern regions, indicating significant lags between outage progression and restoration. Cluster 2 displayed moderate net cumulative customer restoration, with delays concentrated in the western areas. Cluster 0 had the smallest resilience areas, reflecting effective restoration efforts and minimal delays. For the recovery duration, cluster 0 has the longest recovery times, with a consistent median of 17.00 days. This cluster is primarily located in the central and southwestern parts of the region. The uniform shading across the areas within this cluster indicates a consistently delayed recovery process across these regions. Cluster 1 demonstrates shorter recovery durations ranging from 12.00 to 13.50 days, with a median of 12.00 days. This cluster is mainly distributed across the eastern and central regions. The consistent recovery durations across the areas in this cluster suggest an overall efficient and uniform recovery process. Cluster 2 also has a median recovery duration of 12.00 days, identical to cluster 1. The recovery durations in Cluster 2, however, vary more widely, ranging from 11.00 to 31.50 days. Notably, the areas with longer recovery times are predominantly located in the eastern parts of this cluster, indicating localized delays in recovery efforts.

Impact

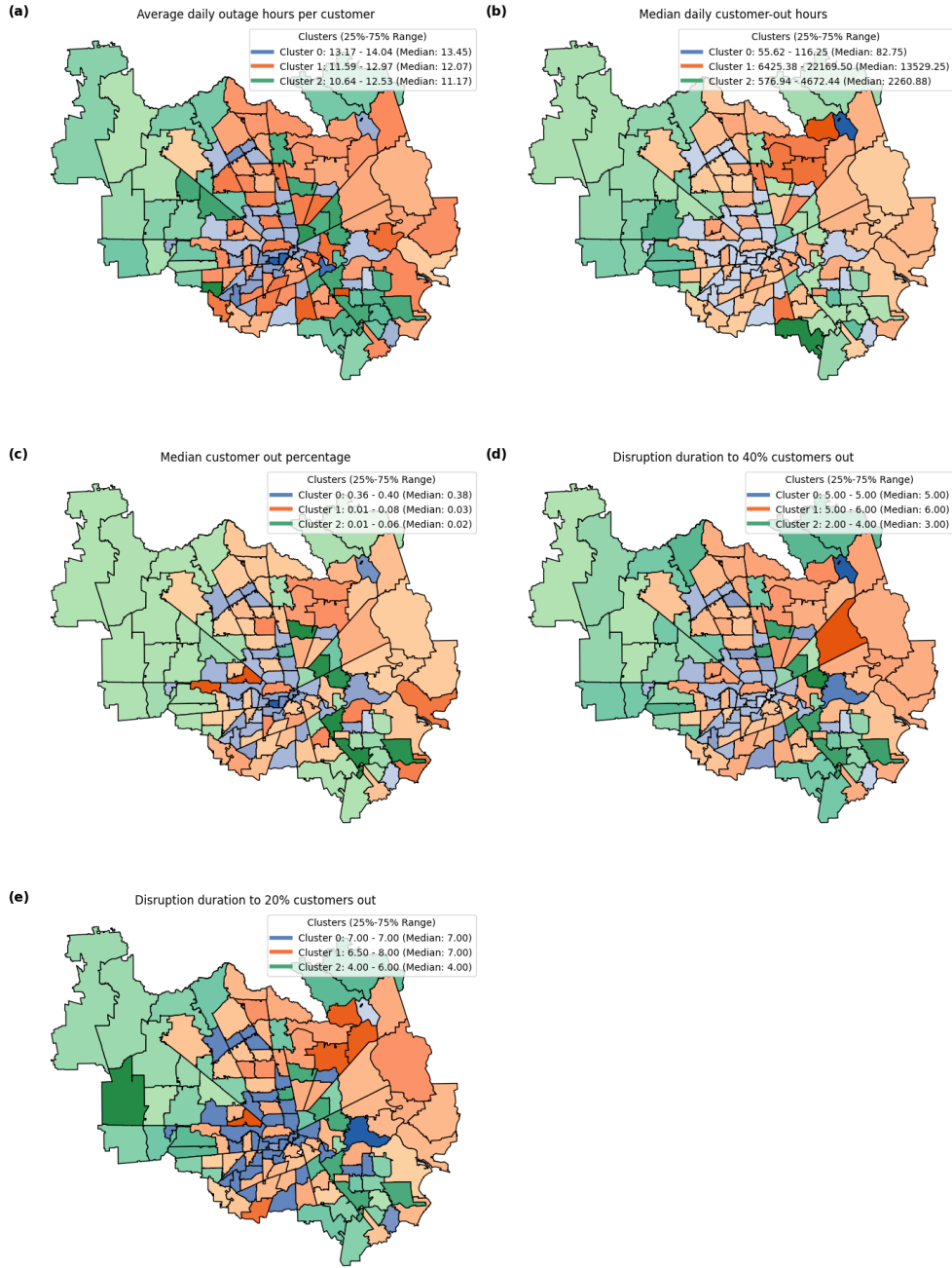

**Figure 14. Spatial distribution of impact features. (a) Average daily outage hours per customer; (b) Median daily customer-out hours; (c) Median customer-out percentage; (d) Disruption duration to 40% customers-out; (e) Disruption duration to 20% customers-out.** Each feature was normalized within its respective cluster to reflect relative values. We used a color gradient to represent low, medium, and high values for each cluster, with shades assigned dynamically

based on ranked cluster medians and interquartile ranges.

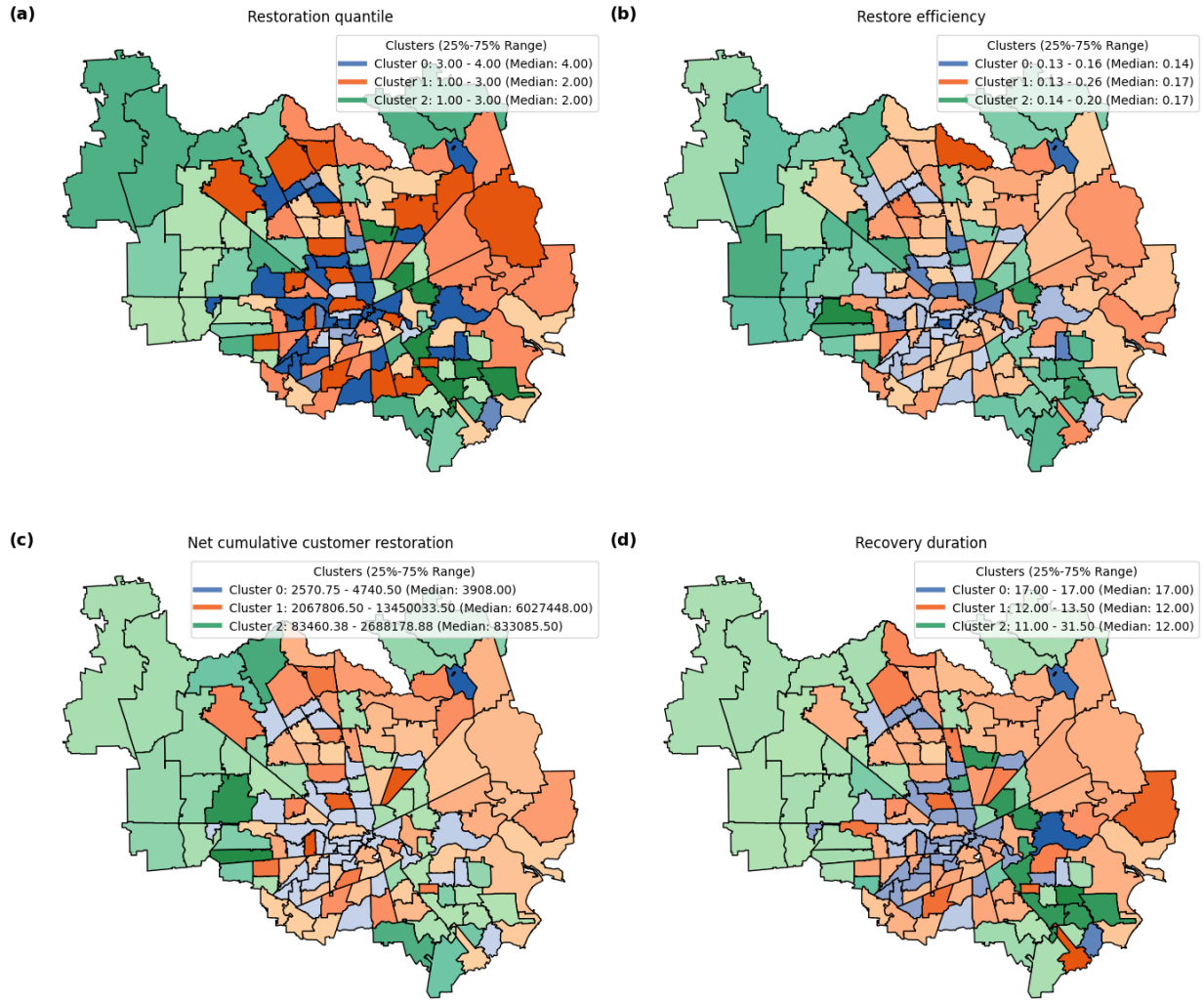

**Figure 14. Spatial distribution of recovery features. (a) Restoration quantile; (b) Restore efficiency; (c) Net cumulative customer restoration; (d) Recovery duration.** Each feature was normalized within its respective cluster to reflect relative values. We used a color gradient to represent low, medium, and high values for each cluster, with shades assigned dynamically based on ranked cluster medians and interquartile ranges.

### 3.4. Determinants of power outage disparity

In order to reveal the underlying factors that shape spatial disparity in power outage impacts and recovery, we analyzed key factors that shape the extent of vulnerability to power outages. Based on the review of the existing literature, we selected the following factors: population density, canopy cover, median and mean income, road density, elevation, proximity to the hurricane path, and whether the region lies within the hurricane cone (Table 2).

**Table 2. Socioeconomic and infrastructural features used in this study.**

| Factor name | Definition | Sources |
|---|---|---|
| Population density | Total population over area | Census Data Reference (Ma and Mostafavi 2024) |
| Canopy | Mean tree canopy percentage in each ZCTA | Multi-resolution Land Characteristics (MRLC) Consortium (U.S. Geological Survey January 14, 2025) |
| Median income | Median household income | Census data Reference (Ma and Mostafavi 2024) |
| Mean income | Mean household income | Census data Reference (Ma and Mostafavi 2024) |
| Road density | Road density in each ZCTA | Reference (Ma and Mostafavi 2024) |
| Elevation | Median elevation in each ZCTA | SRTM 30m Global 1 arc second V003 (Kobrick 2000) |
| Cone | The ZCTA under the direct affected area | KMZ viewer (KML January 13, 2025) |
| Distance to path | Distance to the hurricane path | KMZ viewer (KML January 13, 2025) |

Population density, defined as the total population per unit area, was derived from census data (Ma and Mostafavi 2024). Tree canopy coverage, represented as the mean tree canopy percentage within each ZCTA, was obtained from the Multi-Resolution Land Characteristics (MRLC) Consortium dataset (U.S. Geological Survey January 14, 2025). Sociodemographic indicators, such as median and mean household income, were also sourced from census data (Ma and Mostafavi 2024), providing insights into the economic conditions of the regions analyzed. Road density, calculated as the total length of roads per ZCTA within each ZCTA, was similarly derived from census data (Ma and Mostafavi 2024), reflecting infrastructure characteristics. Elevation data, specifically the median elevation within each ZCTA, was sourced from the SRTM 30m Global 1 arc second V003 dataset (Kobrick 2000), offering a topographic perspective on the regions studied. The cone and distance-to-path was determined using the data from KMZ viewer (KML January 13, 2025), which allows us to view the directly affected areas and the path of the hurricane.

These factors capture a distinct dimension of socioeconomic and infrastructural impact on the hurricane impact and recovery efforts, making them invaluable for explaining the spatial variation in both the extent of power outages and the speed of restoration during hurricanes. Higher population densities not only lead to more stress on aging infrastructure and exacerbate outage impacts but also may warrant prioritization in restoration (Ma and Mostafavi 2024). Proximity to the hurricane's path is a key factor in determining exposure to damaging winds, rainfall, and storm surge—thus intensifying the likelihood of power loss (Snaiki, Wu et al. 2020). Socioeconomic context, represented by median income, further shapes the resources available to communities for infrastructure maintenance and repairs, as well as their ability to advocate for faster service restoration (Ma and Mostafavi 2024). Meanwhile, mean canopy cover can increase the risk of trees and branches damaging power lines, and prolonging outage durations (Ho, Liu et al. 2023). Road density is critical for enabling or hindering repair crews' access to damaged sites, influencing the speed at which power can be restored (Ma and Mostafavi 2024). Also, road network density can capture the complexity of the topology of power grid networks as distribution lines are mainly built along roads (Ma, Li et al. , Rajput, Nayak et al. 2023). Finally, mean elevation not only captures susceptibility to flooding and storm surge in lower-lying areas but also reflects the potential for stronger wind forces at higher elevations (Ho, Lee et al. 2024, Ho, Li et al. 2025).

The boxplots provide a comprehensive visual comparison of key factors across three distinct clusters (Figure 15). Cluster 0 exhibits the highest median population density, suggesting urban areas with concentrated infrastructure and a likely greater demand for rapid restoration. In contrast, clusters 1 and 2, with lower population densities, may represent suburban or rural contexts. Canopy cover varies significantly, with cluster 2 showing the highest median levels, indicating greater vegetation and potentially increased susceptibility to line damage from falling branches, while cluster 0, characterized by lower vegetation, aligns with its urban nature. Income patterns reveal a clear gradient, with Cluster 0 encompassing wealthier neighborhoods potentially benefiting from more robust infrastructure, cluster 2 representing moderate income levels, and cluster 1 reflecting the lowest incomes, potentially limiting resources for preventive measures and restoration. Road density is highest in Cluster 0, gradually decreasing in Clusters 1 and 2, potentially aiding quicker access for repair crews in urban areas. Elevation also varies, with cluster 2 occupying the highest median elevations, possibly reducing flood risks but introducing vulnerabilities like vegetation damage, while clusters 0 and 1 feature lower and intermediate elevations, respectively. Lastly, the clusters differ in their distance to the hurricane path and cone. Cluster 0 exhibits the closest proximity, increasing storm exposure, while Cluster 2 includes areas both far from and directly within the storm's path, reflecting bifurcated risk levels. Synthesizing these insights, Cluster 0 represents higher-income, densely populated, low-elevation areas with well-developed roads and substantial storm exposure; cluster 1 includes lower-income ZIP Codes farther from the storm path but with moderate road density and elevation; and Cluster 2 contains moderately affluent communities with higher elevations and varied storm proximity, balancing reduced flood risks with heightened vegetation damage.

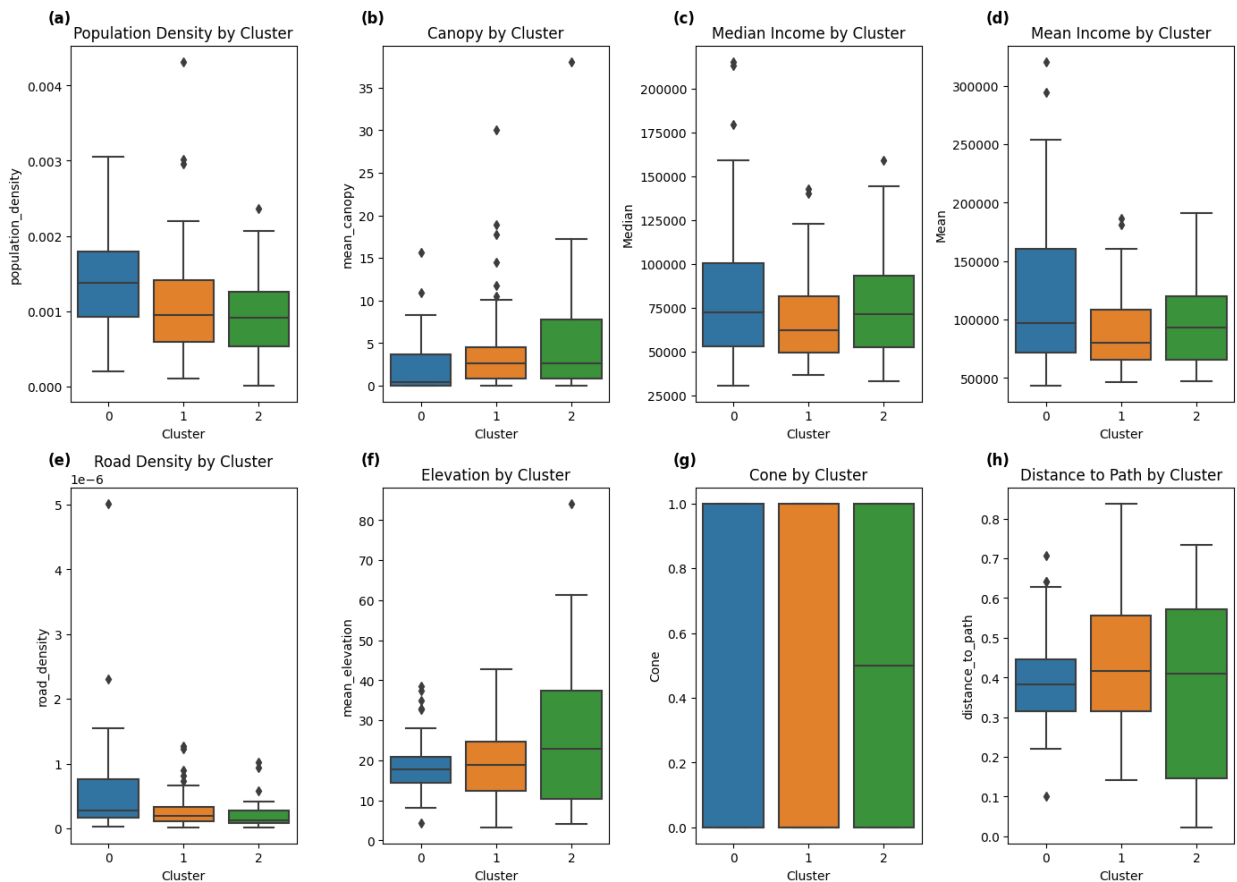

**Fig 15. Box plots of the urban form and structure features across the three clusters. (a) Population Density; (b) Canopy; (c) Median Income; (d) Mean Income; (e) Road Density; (f) Elevation; (g) Cone; (h) Distance to Path**

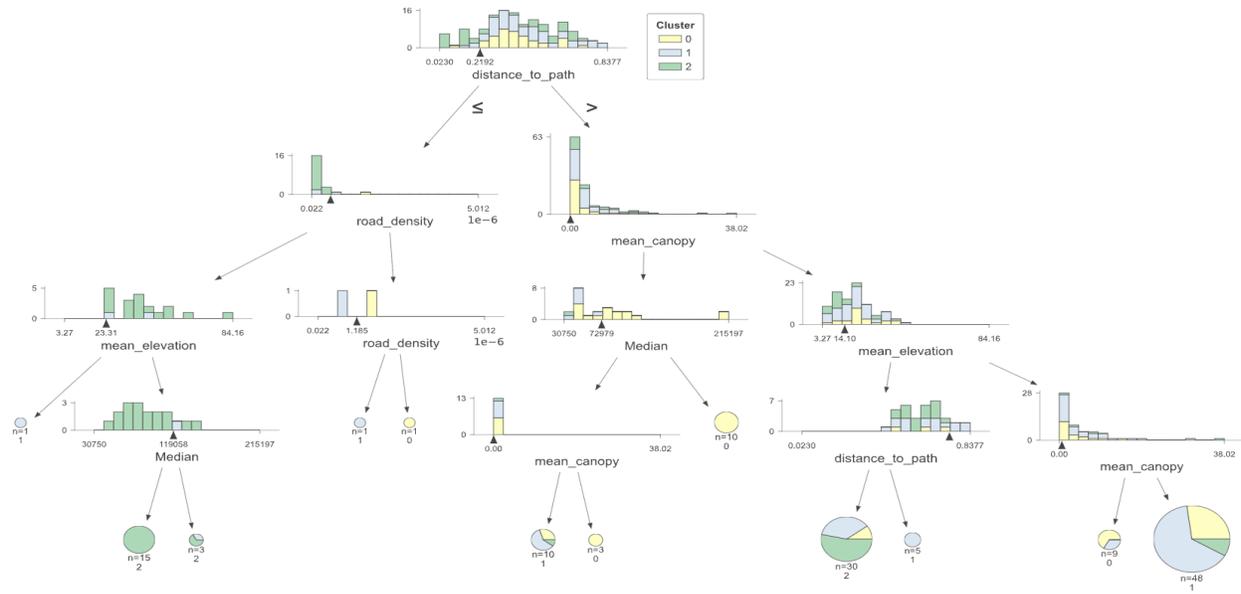

**Fig 16. Decision tree of the socioeconomic and environmental features.** Each node in the tree represents a decision based on a feature and a threshold. The leaf nodes at the end of the tree show the predicted class (cluster) and the number of samples in each class at that node. At each split (node), it displays a histogram showing the distribution of samples for the splitting feature.

To further analyze how the combination of these factors shapes the pathways contributing to different levels of power outage impact and recovery (based on cluster classifications) with socioeconomic and environmental features, we employed a decision tree model (Figure 16). This method was selected to elucidate the hierarchical structure and thresholds of feature values that influence cluster assignments, emphasizing the interpretability of feature interactions. The features analyzed included population density, distance to the hurricane path, median income, mean canopy cover, road density, and mean elevation. These features were chosen to capture the spatial, environmental, and socioeconomic factors that drive community-level impacts and recovery patterns. The decision tree provides a detailed visualization of the pathways that lead to cluster assignments based on the combination of determinant geographic, environmental, and socioeconomic features. The tree consists of 12 distinct pathways, with each pathway representing a unique sequence of feature splits and thresholds that collectively determine the cluster classifications. The root node of the decision tree splits on the distance to the hurricane path, emphasizing its central importance in distinguishing between clusters. Observations with smaller distances are further divided based on features such as road density, mean elevation, and median income. For example, one pathway for smaller distances splits on low road density and high elevation, leading predominantly to cluster 2. Another pathway involves intermediate road density and moderate elevation, resulting in observations being classified into cluster 1. On the other hand, observations with greater distances to the path are split on features like mean canopy cover and road density. These pathways often lead to clusters 0 and 1. For instance, higher canopy values combined with moderate road density typically classify observations into Cluster 0, while lower canopy values tend to lead to Cluster 1. This hierarchical progression highlights how geographic proximity, and environmental features jointly influence cluster assignments.

We also conducted a cluster-wise regression analysis to investigate the relationships between socioeconomic and environmental features with the impact and recovery features. Our aim was to understand the dynamics of power outage impacts and recovery across clusters with distinct socioeconomic and infrastructural characteristics. For each cluster, we ran separate linear regressions to evaluate the effect of predictors on target variables. The use of standardized predictors allowed us to control for variability in units across features. Significant predictors (p-value < 0.05) were identified and interpreted for each cluster, emphasizing their unique roles in shaping outage impacts and recovery outcomes (Table 3).

Our analysis highlighted distinct patterns in clusters 0 and 2, which are characterized by relatively high median incomes. Cluster 0, situated closer to the hurricane path based on its median distance, showed a notable relationship between distance to the path and outage severity. In this cluster, a shorter distance to the path significantly contributed to an increase in impact and recovery metrics, including median daily customer-out hours and net cumulative customer restoration. Moreover, the high population density in cluster 0 is positively associated with the average daily outage hours per customer and median customer out percentage. This finding aligns with the expectation that densely populated areas experience greater challenges during power outages due to the higher demand for resources and complex restoration logistics. A high median income, however, improves their recovery process. Median income emerged as a significant predictor for recovery outcomes in high-income Cluster 0, underscoring the importance of income in enhancing recovery efforts. Specifically, in cluster 0, higher median income was associated with a reduced net cumulative customer restoration, suggesting income inequality in restoration speed across different Zip Codes.

In cluster 2, the distance to the hurricane path exhibited varied effects on both impact and recovery metrics. This cluster includes areas that are either very close to or far from the hurricane's path, resulting in diverse outcomes. For impact metrics, distance to path shows a positive relationship with average daily outage hours per customer (coefficient = 1.046), indicating that areas farther from the hurricane path experienced slightly longer outages on average. The median daily customer-out hours, however, is negatively associated with distance to the path, suggesting that areas closer to the hurricane path experienced significantly more daily outage hours, reflecting the severe nature of the storm's immediate impact on these regions. Population density also has a notable effect, as it is negatively associated with the median customer-out hours (-0.092), implying that higher population density is linked to fewer custom-out hours. Disruption duration to 40% customers shows a positive association with road density (0.82), suggesting that areas with more roads experience longer impact durations, potentially capturing the complexity of the topology of the power grid's downed power lines along the roads. Similarly, days to 20% customer outages is influenced by several factors: it is negatively associated with population density (-1.08), elevation (-1.43), and distance to path (-1.24), indicating that higher-density areas, higher-elevation regions, and areas closer to the path recover faster. However, it is positively associated with road density (0.88), further supporting the notion that denser road networks are linked with the complexity of the topology of power grid networks. Median income also plays a role, with higher-income areas receiving a shorter impact to 20% outages (-1.14), likely due to implicit inequality in restoration prioritization efforts. The restoration quantile, which reflects restoration rates in a seven-day window, shows a negative relationship with population density (-0.63), indicating that higher-density areas exhibit faster restoration rates (suggesting restoration efforts prioritize areas of greater population density). Similarly, the restore efficiency (restore/outage) is negatively associated with elevation (-0.04), suggesting slower restoration in higher-elevation areas. Overall, the findings for cluster 2 highlight the interplay between geographic and socioeconomic factors in shaping both the impact of and recovery from the hurricane. Areas closer to the path bear the brunt of the storm's impact but recover faster, potentially due to prioritized recovery efforts. In contrast, areas farther from the path experience less severe impacts but longer recovery times. The results also suggest the presence of implicit income inequality in prioritization of restoration efforts as areas with higher population had faster restoration.

Cluster 1, with the lowest income levels, faces distinct challenges in mitigating and recovering from power outages. Canopy cover significantly affects recovery times, as tree-related disruptions like fallen branches and obstructed roads slow the restoration process. In Cluster 1, tree canopy is positively associated with disruption duration to 40% customers-out (0.66), suggesting that areas with more tree cover tend to experience more complex damages to power lines and have slightly longer recovery times to reduce outages to 40%. Restoration quantile, a measure of restoration speed in a seven-day window, is positively associated with population density (0.34), indicating that more densely populated areas in this cluster experience slower restoration rates perhaps due to the complexity of damages caused by a higher canopy cover. Population density is associated with slower restoration rate, while higher road density facilitates faster recovery, reflecting a balance between infrastructure strain and repair accessibility. However, low-income levels limit these communities' ability to invest in resilient infrastructure or to mobilize resources, leaving them reliant on external aid. This cluster highlights the vulnerability of resource-constrained areas to prolonged outages, even when direct storm exposure is moderate.

These cluster-specific analyses underscore that hurricane-induced power disruptions result from the interplay of socioeconomic (e.g., income, population density) and infrastructural factors (e.g., elevation, canopy cover, proximity to the path). High-income, densely populated areas (cluster 0) can mitigate some impacts through robust infrastructure, yet direct exposure to storms remains a critical challenge. Mixed-vulnerability areas (cluster 2) require strategies addressing both immediate impacts in high-risk zones and prolonged recovery in outlying areas. Resource-limited communities (cluster 1) would benefit from targeted investments in power grid infrastructure and vegetation management.

Table 3. Linear regression results between predictors (socioeconomic and environmental features) and target variables (power outage impact and recovery features).

| Cluster | Target | Predictor | Coefficient | P-Value |
|---|---|---|---|---|
| 0 | Average daily outage hours per customer | Population density | 0.501 | 0.005 |
| 0 | Median customer out percentage | Population density | 0.046 | 0.006 |
| 0 | Median daily customer-out hours | Distance to path | 1396.820 | 0.012 |
| 0 | Net cumulative customer restoration | Median income | -0.585 | 0.023 |
| 0 | Net cumulative customer restoration | Distance to path | 43378.392 | 0.012 |
| 1 | Disruption duration to 40% customers out | Canopy | 0.656 | 0.024 |
| 1 | Restoration quantile | Population density | 0.337 | 0.018 |
| 1 | Restoration quantile | Road density | -0.258 | 0.025 |
| 1 | Restoration quantile | Median Income | 0.434 | 0.014 |
| 2 | Average daily outage hours per customer | Distance to path | 1.046 | 0.019 |

| 2 | Recovery duration | Population density | -6.689 | 0.031 |
|---|---|---|---|---|
| 2 | Median daily customer-out hours | Population density | -0.092 | 0.001 |
| 2 | Median daily customer-out hours | Distance to path | -2163.479 | 0.034 |
| 2 | Disruption duration to 40% customers out | Road density | 0.818 | 0.001 |
| 2 | Disruption duration to 20% customers out | Population density | -1.086 | 0.004 |
| 2 | Disruption duration to 20% customers out | Road density | 0.880 | 0.004 |
| 2 | Disruption duration to 20% customers out | Elevation | -1.427 | 0.012 |
| 2 | Disruption duration to 20% customers out | Distance to path | -1.241 | 0.029 |
| 2 | Disruption duration to 20% customers out | Median Income | -1.144 | 0.012 |
| 2 | Restoration quantile | Population density | -0.627 | 0.022 |
| 2 | Restore efficiency | Elevation | -0.045 | 0.015 |
| 2 | Restore efficiency | Distance to path | -0.040 | 0.035 |

## 4. Discussion and concluding remarks

Our study provides an extensive spatiotemporal analysis of power outage impacts and recovery patterns following Hurricane Beryl in Houston, Texas. Despite a growing body of research on storm-induced power outages, existing empirical studies remain constrained by their reliance on limited outage metrics and coarse spatial and temporal scales (Hou, Zhu et al. 2021, Dugan, Byles et al. 2023, Flores, Northrop et al. 2024, Zhou, Hu et al. 2024). By focusing on only a narrow range of outage features—often aggregated to large geographic units such as counties—prior work has lacked the granularity to detect nuanced patterns of infrastructure vulnerability and recovery, leaving important intra-community disparities concealed. Also, while many studies highlight spatial inequalities, they frequently overlook the temporal dimension, preventing a robust understanding of how recovery speed varies across communities over time (Coleman, Esmalian et al. 2023, Hsu, Liu et al. 2024, Hsu and Mostafavi 2024). This study addresses these gaps by employing a high-resolution dataset that captures multiple dimensions of power outages—severity, scale, and duration—at the more refined ZIP Code level. In doing so, it not only advances the empirical characterization of widespread outages but also contributes critical insights into the spatiotemporal interplay of sociodemographic, urban development, and environmental factors, thereby offering a comprehensive framework that can inform more equitable and resilient planning and recovery strategies.

The findings of this study offer four important contributions to the state of knowledge and practice. First, the study provides a detailed empirical analysis of power disruptions during a significant hurricane event within a major metropolitan area, yielding valuable insights into urban infrastructure vulnerabilities. Second, the study develops a comprehensive framework for analyzing outage patterns by incorporating multiple metrics, including duration and restoration rates, resulting in a more nuanced understanding of service disruption impacts. The analysis reveals key insights into the dynamics of impact and recovery across the Houston area during Hurricane Beryl. The cumulative outage curve peaked rapidly at approximately $1.3 \times 10^8$ customer power outages (based on 15-minute intervals), reflecting a sudden and widespread disruption. Restoration followed a gradual trajectory, with cumulative restore values converging with the outage curve by July 27, 2024m signaling a return to baseline conditions. The total customer-outage hours amounted to more than 143 million during this period, with the northeastern and central regions experiencing the most severe disruptions, including prolonged average daily outages exceeding 14 hours. Median outage percentages were highest in the northeastern area of the county, reaching up to 84.25%, while central and southeastern areas faced significant disruptions. Recovery metrics highlighted stark disparities: the northeastern areas required up to 15 days to reduce outages to 20%, whereas central and southwestern areas recovered within 5 days. Restoration quantiles and restore efficiency illustrated uneven recovery efforts, with central-eastern areas demonstrating faster restoration rates, while northeastern areas lagged behind. Spatial trends in recovery duration revealed prolonged outages exceeding 40 days in southern and central areas, contrasting with shorter power-outage durations in northern areas. Collectively, these findings emphasize the complex interplay of outage impact, recovery and socioenvironmental factors.

Third, the research harnesses granular outage data at both detailed geographic levels and frequent temporal intervals, revealing previously undetected patterns of spatial inequality. The analysis also identified three distinct clusters with varying vulnerability and recovery profiles. Areas closest to the hurricane path experienced the most severe initial outages but often saw faster restoration, likely due to prioritized response efforts. Cluster 0, characterized by high density and higher income levels, suffered significant initial impacts from direct storm exposure but showed efficient partial recovery rates despite longer total recovery duration. Median income helped reduce impact and recovery Cluster 1, comprising lower-income areas with moderate distance from the storm path and having canopy cover, exhibited slower daily restoration rates, highlighting how limited resources and pre-disaster investments amplify vulnerability. Cluster 2, featuring moderate income levels, variable storm proximity, and higher elevation, showed milder initial impacts but faced extended recovery periods, suggesting potential under-prioritization of peripheral areas. The results highlight significant spatial disparities based on regional characteristics. Areas with high population density and proximity to the hurricane path exhibited the highest impacts, including prolonged outage hours per customer and a higher percentage of customers affected. However, in areas with higher median income, recovery processes were more efficient, as evidenced by lower restoration quantiles, faster recovery durations, and smaller total customer-outage hours, which reflect implicit income inequality in prioritizing restoration efforts.

Fourth, the study conducts a thorough investigation into how various factors—including socioeconomic conditions, urban development patterns, and environmental characteristics—influence the disparities in outage impacts and recovery trajectories across different communities. Several key factors emerged as significant determinants of spatial disparity in outage impacts and recovery rates. Higher median income consistently predicted faster restoration across clusters, underlining the presence of implicit inequality in restoration efforts. Population and road density showed complex effects: while dense urban networks facilitated rapid daily restoration, they also presented logistical challenges that extended complete recovery timeframes. Environmental factors played crucial roles, with elevated areas showing some protection from flooding but occasionally facing slower restoration due to terrain-related challenges. Areas with higher tree canopy experienced more frequent power line damage, leading to increased outage duration. In regions characterized by a mix of proximity to and distance from the hurricane path, recovery outcomes varied. Areas closer to the path experienced more severe initial impacts due to direct exposure but showed faster recoveries, likely due to prioritized recovery efforts. Conversely, areas farther from the path experienced less severe initial impacts but faced prolonged recovery durations. Population density in these regions helped reduce the proportion of affected customers and enhanced restoration rates. However, factors such as road density and elevation occasionally hindered recovery efficiency, highlighting the need for careful consideration of infrastructural complexity. Median income consistently played a significant role in supporting recovery efforts, particularly in areas closer to the hurricane path, underscoring the critical importance of financial resources in disaster resilience. Areas with low median income consistently demonstrated weaker recovery outcomes, regardless of other geographic or environmental advantages. For instance, while road density occasionally facilitated faster restoration by improving accessibility for recovery teams, low-income areas struggled to recover due to constrained resources and limited preparedness. Additionally, environmental features such as tree canopy cover, though generally beneficial, appeared to prolong recovery in some areas by complicating restoration efforts due to downed trees and branches.

Future research would benefit from incorporating higher-resolution grid characteristics data, conducting longitudinal studies of repeated storms, and integrating community engagement to capture local insights into the restoration process. Our findings affirm the critical importance of equity and targeted investments in strengthening both immediate response capabilities and long-term power infrastructure resilience.

**Acknowledgement**

This work was supported by the National Science Foundation under Grant CMMI-1846069 (CAREER). We would also like to thank the Oak Ridge National Laboratory for providing the original power outage data. Any opinions, findings, conclusions, or recommendations expressed in this research are those of the authors and do not necessarily reflect the view of the National Science Foundation and the Oak Ridge National Laboratory.

**Data availability**

The power outage data that support the findings of this study are available pre request from EAGLE-I™. The other datasets used in this paper are publicly accessible and cited in this paper.

**Code availability**

The code that supports the findings of this study is available from the corresponding author upon request.

**Competing interests**

The authors declare no competing interests.